\documentclass{article}
\usepackage{graphicx}

\PassOptionsToPackage{numbers, compress}{natbib}

\usepackage[final]{neurips_2024}

\usepackage[utf8]{inputenc} 
\usepackage[T1]{fontenc}    
\usepackage{hyperref}       
\usepackage{url}            
\usepackage{booktabs}       
\usepackage{amsfonts}       
\usepackage{nicefrac}       
\usepackage{microtype}      
\usepackage{xcolor}         

\title{Position: Open and Closed Large Language Models in Healthcare}

\author{%
  Jiawei Xu \\
  School of Information\\
  University of Texas at Austin\\
  \texttt{jiaweixu@utexas.edu} \\
  \And
  Ying Ding \\
  School of Information \\
  University of Texas at Austin \\
  \texttt{ying.ding@ischool.utexas.edu} \\
  \AND
  Yi Bu \\
  Department of Information Management \\
  Peking University \\
  \texttt{buyi@pku.edu.cn} \\
}

\begin{document}

\maketitle

\begin{abstract}
This position paper provides an analysis of the evolving roles of open-source and closed-source large language models (LLMs) in healthcare, emphasizing their distinct contributions and the scientific community's response to their development. Closed LLMs, such as GPT-4, have dominated high-performance applications, particularly in medical imaging and multimodal diagnostics, due to their advanced reasoning capabilities. Conversely, open LLMs, like Meta’s LLaMA, have gained popularity for their adaptability and cost-effectiveness, enabling researchers to fine-tune models for specific domains, such as mental health and patient communication.
\end{abstract}

\section{Background}

Large language models represent a significant scientific advancement. In the field of large language models, OpenAI’s GPT-4~\cite{Achiam2023GPT4TR} has the highest performance since its release. However, it is considered a “closed-source” or “black-box” model, as it does not release its model weights to the public and access to these weights is restricted under proprietary licenses. Access to GPT-4 is provided through platforms such as the web app ChatGPT.com or the OpenAI API. Conversely, “open-source” LLMs typically offer downloadable model weights and are governed by non-proprietary licenses, exemplified by Meta's LLaMA 2~\cite{Touvron2023Llama2O}. The model weights represent the distilled knowledge that an LLM acquires from extensive text datasets. Open-source LLMs provide access to model weights, thus enabling adaptation (continuing pretraining and fine-tuning) and facilitating further investigation of the model on local devices \cite{solaiman2023gradient}. For instance, researchers can deploy LLaMA 2 on their local devices using its model weights. More importantly, they can further pretrain or fine-tune LLaMA 2 with their datasets at a much lower cost than training a model from scratch. One notable example is Princeton University’s Llemma~\cite{Azerbayev2023LlemmaAO}, an LLM for mathematics that continues pretraining on LLaMA 2 using the mathematical corpus Proof-Pile-2, demonstrating significant mathematical abilities. While closed LLMs lead in performance, open-source LLMs are distinguished by their adaptability.

\section{Data and Methods}
Our data consists of two parts: (1) {\bf data about teams training LLMs} and (2) {\bf metadata of scientific publications mentioning LLM}. Both segments distinguish between open LLMs and closed LLMs.

\subsection{Teams Training LLMs}
Model-related data provides a clear illustration of the evolution of LLMs, from the early GPT-2 to the recent LLaMA 3. We derived this data from the Ecosystem Graph for Foundation Models \cite{Bommasani2023EcosystemGT}. This graph categorizes the access type (either ‘open’ or ‘closed’) of foundation models, which include both LLMs and other generative AI models. By utilizing the GPT-4o API and conducting manual reviews, we examined all listed foundation models as of April 2024, identifying a total of 201 LLMs, of which 54 are closed and 147 are open. We located the original papers for 100 open-source LLMs and 34 closed-source LLMs, allowing us to analyze the team structures behind these models. Among the 115 organizations that developed these 201 LLMs, 62 are companies, 29 are universities, 12 are government-funded or non-profit research institutes, and 12 are global open initiatives or projects.
\subsection{Teams Using, Evaluating, or Mentioning LLMs}
Using official arXiv data accessed via the Kaggle API, with a cutoff date of June 1, 2024 (2.94 million papers), we selected papers that mention specific LLM model names in their titles and abstracts. This allowed us to explore papers related to both open-source and closed-source LLMs. Our two-stage retrieval strategy identified papers associated with open-source and closed-source models. In the first stage, we searched for general LLM-related papers within the 2.94 million papers in arXiv dataset, yielding 30,111 papers. In the second stage, we identified papers mentioning open-source and closed-source LLMs from the initial set, matching 44 closed-source LLMs and 103 open-source LLMs within the titles and abstracts. This process identified a total of 6,198 papers mentioning either open-source or closed-source specific LLMs in the titles and abstracts. Descriptive statistics for the LLM-related papers are provided in Table~\ref{table-1}.

\begin{table}[ht]
  \caption{Summary of papers mentioning specific LLMs in titles and abstracts}
  \label{table-1}
  \centering
  \begin{tabular}{lll}
    \toprule
    \cmidrule(r){1-2}
    Statistic Description     & Number \\
    \midrule
    Open Source LLMs (matched) & 147 (103) \\
    Closed Sourced LLMs (matched) & 54 (44)   \\
    Papers mentioning specific closed-source LLMs & 2,982 \\
    Papers mentioning specific open-source LLMs & 2,283   \\
    Papers mentioning both closed-source LLMs and open-source LLMs & 933 \\
    Total papers mentioning specific LLMs & 6,198   \\
    \bottomrule
  \end{tabular}
\end{table}

\section{Results}
\begin{figure}[ht]
  \centering
  \includegraphics[width=.95\linewidth]{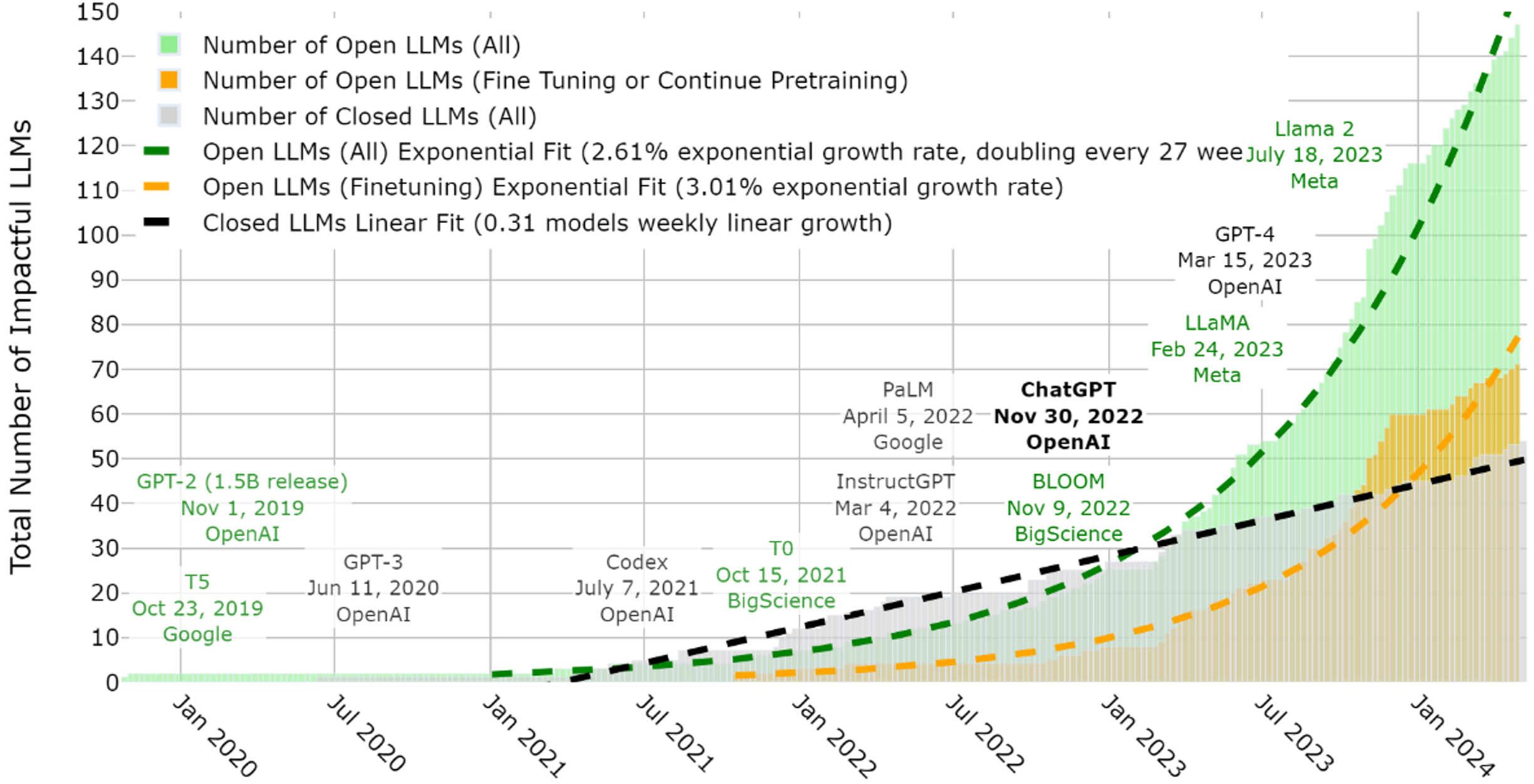}
  \caption{Growth rates of open and closed LLMs.}
  \label{figure-1}
\end{figure}

We analyzed 201 influential large language models (LLMs) identified by the Stanford Center for Research on Foundation Models (CRFM) as of April 30, 2024, and examined the 6,198 arXiv papers that mention these LLMs in their titles and abstracts. Per Figure~\ref{figure-1}, our investigation yielded two major findings: (1) Elite organizations, such as major technology companies and startups, predominantly drive LLM training, whether for closed or open models. We observed that the number of open LLMs grows exponentially, whereas the number of closed LLMs increases linearly. This exponential growth of open LLMs is likely due to their model weight accessibility, allowing the public to fine-tune and continue pre-training on existing models, thereby reducing the significant costs associated with training an LLM from scratch. This democratizes the process, enabling smaller, less experienced, and non-industrial teams to train domain-specific LLMs at a lower cost. (2) The scientific community’s reaction to closed and open LLMs varies significantly. Overall, although both open and closed models are experiencing increased attention, closed LLMs have garnered more scientific interest since 2022. Specifically, closed LLMs are notably more popular in fields such as Computers and Society (cs.CY) and Human-Computer Interaction (cs.HC), where researchers prioritize leveraging the LLMs’ powerful reasoning abilities over the training process itself. Conversely, in domains like Machine Learning (cs.LG) and Distributed, Parallel, and Cluster Computing (cs.DC), open-weight LLMs are preferred due to researchers’ focus on optimizing training processes and model adaptation.

Model creators can choose to keep LLM weights proprietary or open them to the public. Analyzing the total number of closed and open models from 2019 to 2024, we noted that closed LLMs (represented by gray bars in Figure 1) exhibit linear growth. In contrast, the number of open LLMs (green bars) has grown exponentially, with those fine-tuned on existing models (orange bars) contributing significantly to this trend. The growth rate of closed models initially outpaced open models until 2023, after which open models surged, primarily driven by fine-tuning efforts. Closed LLMs gained prominence in June 2020 with the release of GPT-3~\cite{Brown2020LanguageMA} by OpenAI. With its massive size (175 B parameters), impressive few-shot learning capabilities, and the heavy investment behind it, OpenAI opted to keep GPT-3’s model weights proprietary and prevent the public from continuing training or finetuning on it with existing model weights, offering access only through an official API. This approach was subsequently adopted by other companies, including Microsoft and Nvidia with their Megatron-Turing NLG~\cite{Smith2022UsingDA} (530 B parameters, Jan 2022) and Google’s PaLM~\cite{Chowdhery2022PaLMSL} (540 B parameters, April 2022), etc. Since then, the growth of closed-source LLMs has been steady with a linear trajectory.

In contrast, early open LLMs like Google’s T5~\cite{Raffel2019ExploringTL} and OpenAI’s GPT-2~\cite{Radford2019LanguageMA} in 2019 adhered to the open-source tradition, releasing their model weights (more than 1 billion parameters) along with the paper so that the public can continue training or finetuning their models for their own use or research. However, following the closed release of GPT-3 in 2020, the most powerful and best-performance LLMs all choose to keep proprietary. Open LLM communities were relatively non-prominent during this period. Meta was the first major tech company to disrupt the closed-source paradigm with the release of its LLaMA model~\cite{Touvron2023LLaMAOA} in February 2023, which was comparable to OpenAI’s GPT-3.5 at that time. This event catalyzed a surge in open LLMs. Notably, many new open models, such as LMSYS’s Vicuna~\cite{chiang2023vicuna}, Stanford’s Alpaca~\cite{taori2023alpaca}, and UC Berkeley’s OpenLLaMA, were fine-tuned versions of LLaMA. Meta’s decision to open source the LLaMA series has contributed to the exponential growth of open LLMs, with other companies like Mistral and Alibaba following Meta’s open source strategy. This strategy involves keeping the model weights open to the public while also providing API and web applications. The exponential growth trend continues, as state-of-the-art open LLMs have increasingly closed the gap with their closed counterparts. Fine-tuning has proven to be cost-effective and allows for more specialized models in specific domains. 

The scientific community’s attention to LLMs varies, with prominent models like GPT-3.5, GPT-4, and LLaMA receiving significant recognition, while less notable LLMs attract limited focus. We analyzed the trend of scientific attention towards closed LLMs (gray line in Figure 2) and open LLMs (green line) from 2019 to 2024, using mentions in arXiv papers as a proxy. BERT~\cite{Devlin2019BERTPO}, an important language model before the LLM era, is included as a baseline (dashed line). By 2024, closed LLMs, led by the GPT-4 series, had garnered more scientific attention than open LLMs. Scientific interest in both types surged after the release of ChatGPT in late 2022, but the rise was more pronounced for closed LLMs like GPT-3.5 and GPT-4, owing to their superior performance and accessibility through web interfaces and APIs.

\begin{figure}[ht]
  \centering
  \includegraphics[width=.9\linewidth]{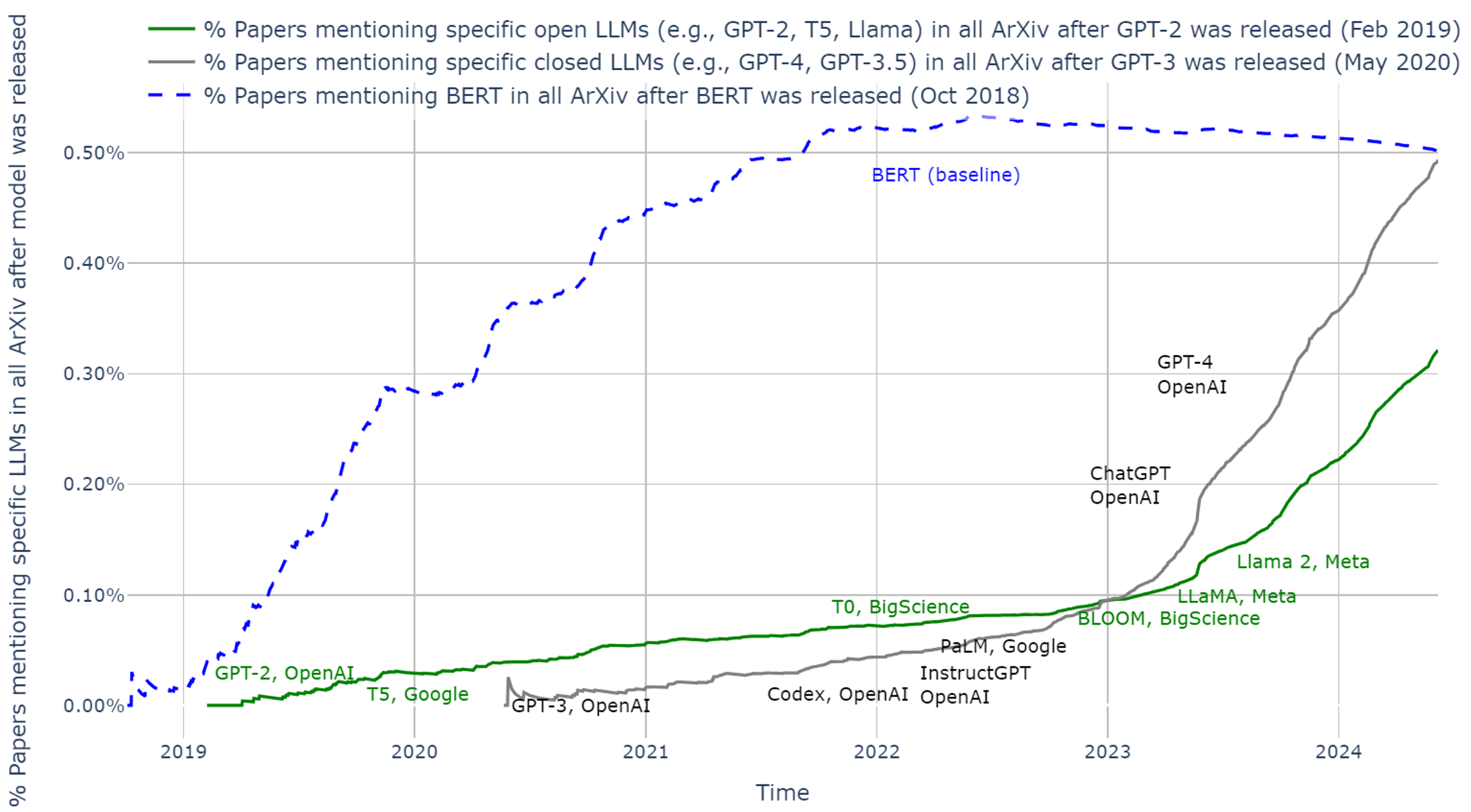}
  \caption{Percentage of arXiv papers mentioning open LLMs or closed LLMs from 2019 onwards, with BERT as a baseline.}
  \label{figure-2}
\end{figure}

Open LLMs such as GPT-2 and T5 initially received considerable attention. Their open weights allowed researchers to fine-tune and continue training to explore their capabilities. However, this shifted with the release of ChatGPT, which, despite being closed, demonstrated superior natural language understanding and usability, attracting extensive scientific interest. State-of-the-art performance by closed LLMs, notably OpenAI’s models, has made them the preferred choice for tasks requiring top-tier capabilities.

\begin{figure}[ht]
  \centering
  \includegraphics[width=1.0\linewidth]{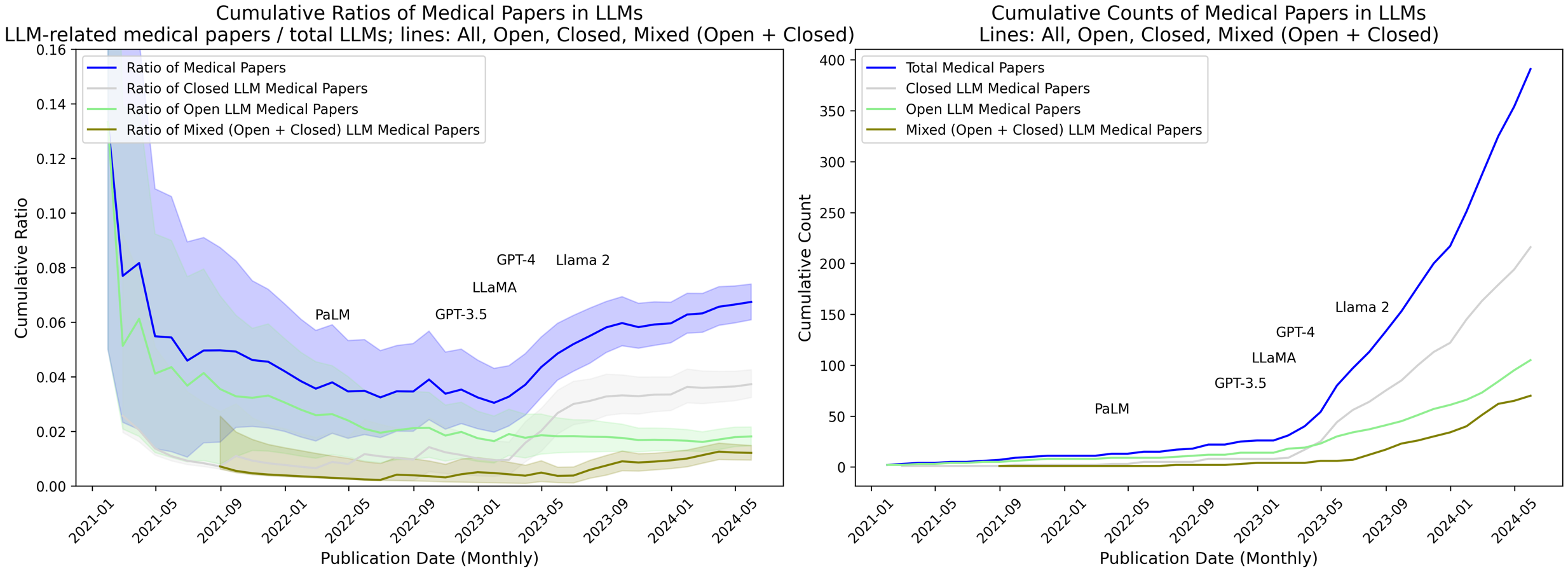}
  \caption{Cumulative ratio (left) and counts (right) of medical papers in LLMs.}
  \label{figure-3}
\end{figure}

Among these LLM papers, we obtain 404 publications from the domain of medical sciences. Figure~\ref{figure-3} highlights the increasing prominence of LLMs in the field of healthcare. The left part of the figure displays the cumulative ratio of medical-related LLM publications, showing a steady rise in the interest of the scientific community in applying both open-source and closed-source LLMs to healthcare-related domains. This trend became particularly pronounced starting in 2023 when healthcare professionals and AI researchers began leveraging these models for applications ranging from diagnostic imaging to predictive analytics and conversational agents for patient care.
On the right side of Figure~\ref{figure-3}, we observe the cumulative number of medical papers mentioning LLMs. The data suggest a clear shift toward the adoption of closed-source models, particularly due to their state-of-the-art performance in complex medical tasks such as radiology imaging and multimodal medical reporting. Closed models like GPT-4 are gaining traction for their advanced capabilities, which have proven beneficial in handling sensitive patient data in clinical environments. However, open-source models are also becoming increasingly relevant, particularly in areas like mental health and patient communication, where fine-tuning on specific datasets allows for more personalized and domain-specific applications. This figure underscores the growing importance of both types of LLMs in healthcare, with closed models dominating high-performance tasks and open models enabling specialized, cost-effective solutions in niche areas.

\begin{figure}[ht]
  \centering
  \includegraphics[width=1\linewidth]{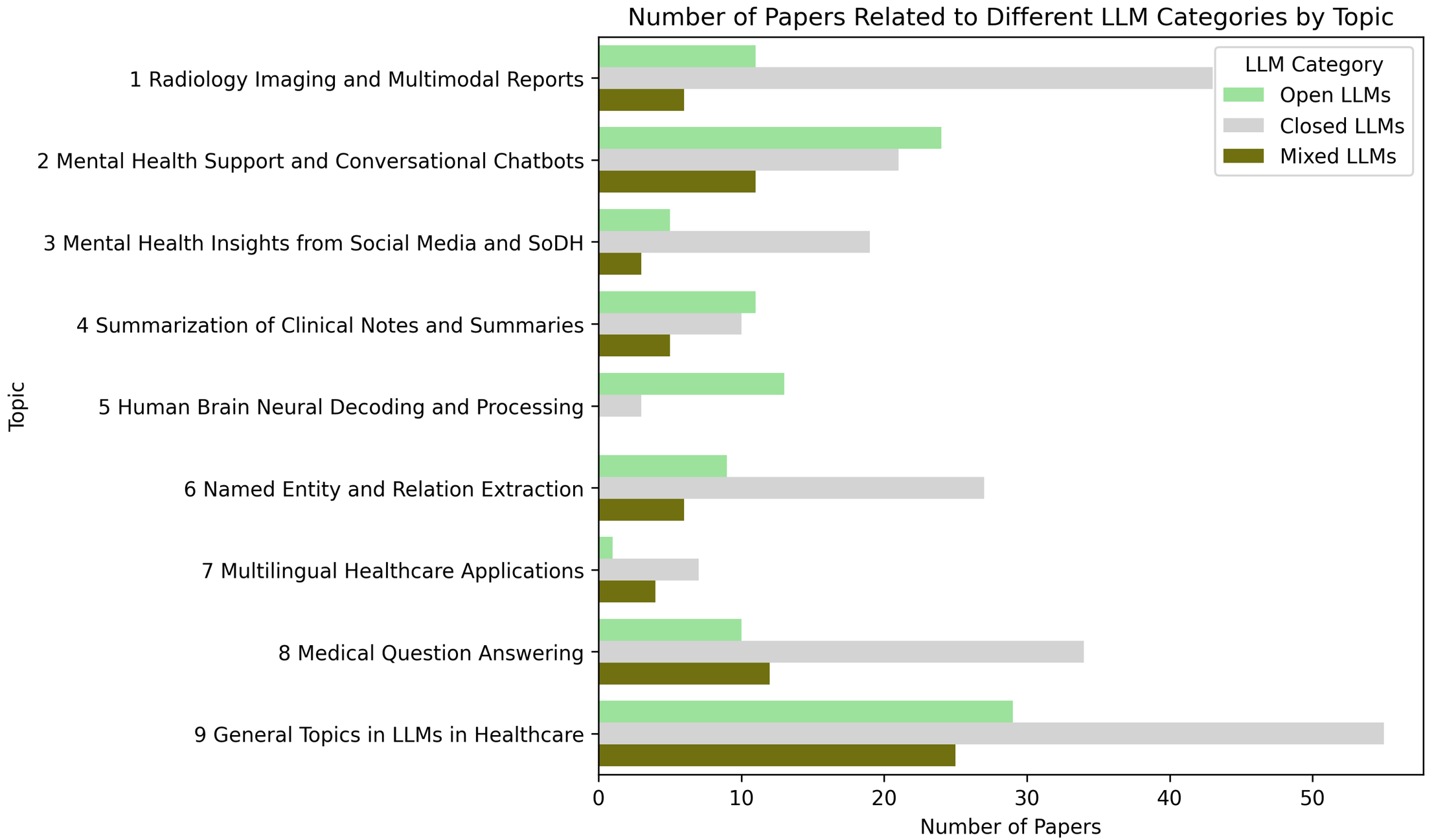}
  \caption{Topic modeling results for medical LLM papers.}
  \label{figure-4}
\end{figure}

We further adopt the BERTopic~\cite{grootendorst2022bertopic} model for topic modeling of all papers mentioning specific open or closed LLM names. Figure~\ref{figure-4} presents a topic modeling analysis of medical-related LLM papers, offering insights into the distinct research directions driven by open and closed LLMs in the healthcare sector. We see that closed LLM papers focus predominantly on high-complexity tasks such as radiology and medical imaging, where precision and reasoning capabilities are paramount. These models' superior performance in these areas makes them a go-to solution for research requiring high accuracy and reliability in medical diagnostics.
On the other hand, papers related to open-source LLMs show a different trend, with a higher concentration of topics focused on mental health applications, conversational agents, and support systems in healthcare. Open LLMs’ flexibility allows researchers to fine-tune models on specific healthcare-related datasets, creating personalized and context-aware solutions, particularly in mental health support systems and patient-doctor communication.

These topic modeling results emphasize the complementary roles of open and closed LLMs in the healthcare landscape. Closed models are spearheading advancements in high-stakes medical applications, while open models excel in areas that benefit from their adaptability and fine-tuning capabilities. Moving forward, this divergence in focus highlights the potential for hybrid approaches, combining the strengths of both open and closed models to address the diverse needs of healthcare systems globally.

\section{Summary}
This position paper explores the rapid development of LLMs in healthcare, particularly focusing on the differences between open-source and closed-source models. Closed models lead in high-performance applications like radiology and medical imaging, where their superior reasoning capabilities are highly valued. These models have become prominent in healthcare because they provide powerful tools for complex diagnostic tasks, making them indispensable for research and clinical applications that require top-tier performance. On the other hand, open LLMs, exemplified by models such as Meta’s LLaMA, are democratizing AI research in healthcare by allowing researchers to fine-tune them for specific applications, such as mental health support and conversational agents. This adaptability makes open LLMs particularly useful in fields where customization is important, and their lower costs make them accessible to smaller research teams and institutions. 

\section*{Acknowledgments}
Yi Bu acknowledges the financial support of the National Natural Science Foundation of China (\#72104007 and \#72174016).

\bibliographystyle{plain}
\bibliography{references.bib}

\begin{thebibliography}{10}

\bibitem{Touvron2023Llama2O}
Meta AI.
\newblock Llama 2: Open foundation and fine-tuned chat models.
\newblock {\em ArXiv}, abs/2307.09288, 2023.

\bibitem{Azerbayev2023LlemmaAO}
Zhangir Azerbayev, Hailey Schoelkopf, Keiran Paster, Marco~Dos Santos, Stephen~Marcus McAleer, Albert~Q. Jiang, Jia Deng, Stella Biderman, and Sean Welleck.
\newblock Llemma: An open language model for mathematics.
\newblock {\em ArXiv}, abs/2310.10631, 2023.

\bibitem{Bommasani2023EcosystemGT}
Rishi Bommasani, Dilara Soylu, Thomas Liao, Kathleen~A. Creel, and Percy Liang.
\newblock Ecosystem graphs: The social footprint of foundation models.
\newblock {\em ArXiv}, abs/2303.15772, 2023.

\bibitem{Brown2020LanguageMA}
Tom~B. Brown, Benjamin Mann, Nick Ryder, Melanie Subbiah, Jared Kaplan, Prafulla Dhariwal, Arvind Neelakantan, Pranav Shyam, Girish Sastry, Amanda Askell, Sandhini Agarwal, Ariel Herbert-Voss, Gretchen Krueger, Tom Henighan, Rewon Child, Aditya Ramesh, Daniel~M. Ziegler, Jeff Wu, Clemens Winter, Christopher Hesse, Mark Chen, Eric Sigler, Ma~teusz Litwin, Scott Gray, Benjamin Chess, Jack Clark, Christopher Berner, Sam McCandlish, Alec Radford, Ilya Sutskever, and Dario Amodei.
\newblock Language models are few-shot learners.
\newblock {\em ArXiv}, abs/2005.14165, 2020.

\bibitem{chiang2023vicuna}
Wei-Lin Chiang, Zhuohan Li, Zi~Lin, Ying Sheng, Zhanghao Wu, Hao Zhang, Lianmin Zheng, Siyuan Zhuang, Yonghao Zhuang, Joseph~E Gonzalez, et~al.
\newblock Vicuna: An open-source chatbot impressing gpt-4 with 90\%* chatgpt quality, march 2023.
\newblock {\em URL https://lmsys. org/blog/2023-03-30-vicuna}, 3(5), 2023.

\bibitem{Devlin2019BERTPO}
Jacob Devlin, Ming-Wei Chang, Kenton Lee, and Kristina Toutanova.
\newblock Bert: Pre-training of deep bidirectional transformers for language understanding.
\newblock In {\em North American Chapter of the Association for Computational Linguistics}, 2019.

\bibitem{Chowdhery2022PaLMSL}
Google.
\newblock Palm: Scaling language modeling with pathways.
\newblock {\em J. Mach. Learn. Res.}, 24:240:1--240:113, 2022.

\bibitem{grootendorst2022bertopic}
Maarten Grootendorst.
\newblock Bertopic: Neural topic modeling with a class-based tf-idf procedure.
\newblock {\em arXiv preprint arXiv:2203.05794}, 2022.

\bibitem{Achiam2023GPT4TR}
OpenAI.
\newblock Gpt-4 technical report.
\newblock {\em ArXiv}, 2023.

\bibitem{Radford2019LanguageMA}
Alec Radford, Jeff Wu, Rewon Child, David Luan, Dario Amodei, and Ilya Sutskever.
\newblock Language models are unsupervised multitask learners.
\newblock {\em ArXiv}, 2019.

\bibitem{Raffel2019ExploringTL}
Colin Raffel, Noam~M. Shazeer, Adam Roberts, Katherine Lee, Sharan Narang, Michael Matena, Yanqi Zhou, Wei Li, and Peter~J. Liu.
\newblock Exploring the limits of transfer learning with a unified text-to-text transformer.
\newblock {\em J. Mach. Learn. Res.}, 21:140:1--140:67, 2019.

\bibitem{Smith2022UsingDA}
Shaden Smith, Mostofa Patwary, Brandon Norick, Patrick LeGresley, Samyam Rajbhandari, Jared Casper, Zhun Liu, Shrimai Prabhumoye, George Zerveas, Vijay~Anand Korthikanti, Elton Zhang, Rewon Child, Reza~Yazdani Aminabadi, Julie Bernauer, Xia Song, Mohammad Shoeybi, Yuxiong He, Michael Houston, Saurabh Tiwary, and Bryan Catanzaro.
\newblock Using deepspeed and megatron to train megatron-turing nlg 530b, a large-scale generative language model.
\newblock {\em ArXiv}, abs/2201.11990, 2022.

\bibitem{solaiman2023gradient}
Irene Solaiman.
\newblock The gradient of generative ai release: Methods and considerations.
\newblock In {\em Proceedings of the 2023 ACM Conference on Fairness, Accountability, and Transparency}, FAccT '23, pages 111--122. Association for Computing Machinery, 2023.

\bibitem{taori2023alpaca}
Rohan Taori, Ishaan Gulrajani, Tianyi Zhang, Yann Dubois, Xuechen Li, Carlos Guestrin, Percy Liang, and Tatsunori~B Hashimoto.
\newblock Alpaca: A strong, replicable instruction-following model.
\newblock {\em Stanford Center for Research on Foundation Models. https://crfm. stanford. edu/2023/03/13/alpaca. html}, 3(6):7, 2023.

\bibitem{Touvron2023LLaMAOA}
Hugo Touvron, Thibaut Lavril, Gautier Izacard, Xavier Martinet, Marie-Anne Lachaux, Timoth{\'e}e Lacroix, Baptiste Rozi{\`e}re, Naman Goyal, Eric Hambro, Faisal Azhar, Aurelien Rodriguez, Armand Joulin, Edouard Grave, and Guillaume Lample.
\newblock Llama: Open and efficient foundation language models.
\newblock {\em ArXiv}, abs/2302.13971, 2023.

\end{thebibliography}

\end{document}